\documentclass[aps,prb,superscriptaddress,twocolumn,showpacs]{revtex4-1}
\usepackage{amsmath,amssymb,mathrsfs,bbold}
\usepackage[pdftex]{graphicx}
\usepackage{epstopdf}
\newcommand{\comm}[2]{\left[#1,#2\right]}

\newcommand{\ket}[1]{\left|#1\right\rangle}

\newcommand{\average}[1]{\langle#1\rangle}
\newcommand{\abs}[1]{{\lvert{#1}\rvert}}

\newcommand{\logTerm}[2]{\mathcal L_{#1}(#2)}

\DeclareMathOperator{\tr}{Tr}

\allowdisplaybreaks[4]

\begin{document}
%%%%%%%%%%%%%%%%%%%%%%%%%%%%%%%%%%%%%%%%%%%%%%
\title{Renormalization-group analysis of a spin-1 Kondo dot:\\ 
Non-equilibrium transport and relaxation dynamics}
%%%%%%%%%%%%%%%%%%%%%%%%%%%%%%%%%%%%%%%%%%%%%%
\author{Christoph B. M. H\"orig}
\author{Dirk Schuricht}
\author{Sabine Andergassen}
\affiliation{Institute for Theory of Statistical Physics, 
RWTH Aachen, 52056 Aachen, Germany} 
\affiliation{JARA-Fundamentals of Future Information Technology}
 \date{\today}
\pagestyle{plain}

\begin{abstract}
We study non-equilibrium transport through a spin-1 Kondo dot in
a local magnetic field. To this end we perform a two-loop
renormalization group analysis in the weak-coupling regime yielding analytic results for (i) the renormalized 
magnetic field and the g-factor, (ii) the time evolution of observables and the relevant 
decay rates, (iii) the magnetization and anisotropy as well as (iv) the current and 
differential conductance in the stationary state. In particular, we find that compared to a
spin-1/2 Kondo dot there exist three additional decay rates resulting in an enhanced
broadening of the logarithmic features observed in stationary quantities. Additionally, we study the effect of anisotropic couplings between reservoir and impurity spin.
\end{abstract}
%05.10.Cc	Renormalization group methods
%73.63.Kv	Quantum dots
%72.10.Bg	General formulation of transport theory

\maketitle
%%%%%%%%%%%%%%%%%%%%%%%%%%%%
\section{Introduction}
%%%%%%%%%%%%%%%%%%%%%%%%%%%%
The Kondo model consists of a localized spin-1/2 which is coupled to the
electrons in the host metal via a spin exchange interaction. The original
motivation to study~\cite{Kondo64,Hewson93} the properties of this and
related models was the observation of a resistance minimum in some metals 
in the 1930s. More than two decades ago it was 
realized~\cite{GlazmanRaikh88,NgLee88,Kawabata91} that the Kondo model 
can also be applied to describe transport experiments through quantum
dots, where the localized spin-1/2 models a minimal two-state quantum dot 
which is coupled via exchange interactions to two (or more) leads held at
different chemical potentials.
The tremendous developments in the ability to engineer nanoscale 
devices also led to the experimental observation~\cite{Goldhaber-98,Cronenwett-98,Schmid-98,Goldhaber-98prl,Simmel-99,Wiel-00,Nygard-00} 
of Kondo physics in transport experiments.
This in turn triggered many theoretical 
studies~\cite{SchoellerKoenig00,Rosch-01,ParcolletHooley02,Rosch-03prl,Paaske-04prb1,Paaske-04prb2,Rosch-05,Kehrein05,Kehreinbook,DoyonAndrei06,Korb-07,Anders08prl,SchoellerReininghaus09,SS09,FritschKehrein10,PS11} 
of the non-equilibrium transport properties of Kondo quantum dots.

It is well-known that the Kondo model with larger spin on the dot 
shows~\cite{NozieresBlandin80,Andrei-83,TsvelikWiegmann83,Hewson93} 
strong deviations from standard Fermi liquid behavior at low temperatures. For example, 
at low temperatures the local spin is not fully screened by the electrons
and a residual magnetization remains. Despite the recent 
observation of this underscreened Kondo effect in transport experiments
in single molecules~\cite{Roch-08,Roch-09,Parks-10,Florens-11}, the non-equilibrium
transport properties of the spin-1 Kondo model have not been studied 
theoretically~\cite{PosazhennikovaColeman05,Paaske-06,Posazhennikova-07,RouraBasAligia09,RouraBasAligia10,Florens-11,Wright-11} 
in the same detail as its spin-1/2 counterpart.

In this article we fill this gap by performing a two-loop renormalization group (RG) analysis 
for a spin-1 Kondo dot with finite bias voltage. To this end we apply the real-time 
renormalization group method in frequency space~\cite{Schoeller09} (RTRG-FS)
to derive analytic results for the renormalized magnetic field and g-factor, decay
rates and time evolution as well as magnetization, anisotropy and differential 
conductance in the stationary state. All results are obtained in a systematic weak-coupling 
expansion in the renormalized exchange coupling $J(\Lambda)\ll1$. The latter condition
is satisfied as long as the maximum of all occurring energy scales is much larger than the Kondo temperature $T_K$,
\begin{equation}
\Lambda_c=\max\lbrace V,h_0\rbrace\gg T_K,
\label{eq:wcc}
\end{equation}
where $V$ and $h_0$ are the applied bias voltage and magnetic field, respectively, and
$T_K$ denotes the Kondo temperature at which the system enters the strong-coupling
regime. We find that the properties of the spin-1 
Kondo model are similar to the spin-1/2 system with the main differences (i) that
the larger local Hilbert space on the dot results in additional (quintet) decay rates
beside the usual longitudinal and transverse spin relaxation rates, (ii) that these 
rates describe the decay of the spin anisotropy $T\propto S^iS^j$ (which is
trivial in the spin-1/2 model), where $S^i$ are the components of the spin operator
on the dot, and (iii) that the existence of these additional rates yields a broadening 
(relative to the spin-1/2 model) of logarithmic features in stationary quantities like 
the susceptibilities and the differential conductance.

The paper is structured as follows. In the next two sections we define the spin-1 Kondo dot 
and briefly review the applied RTRG-FS method. In Sec.~\ref{sec:results} we present
the results for the renormalized magnetic field and g-factor, decay rates and time 
evolution as well as magnetization, anisotropy and current in the stationary state. 
In particular, we highlight the differences to the spin-1/2 model studied in 
Ref.~\onlinecite{SchoellerReininghaus09}. In Sec.~\ref{sec:anisotropic-model} we investigate
the effect of anisotropic exchange couplings. 
Our calculations mainly follow the analysis of the spin-1/2 Kondo 
model~\cite{SchoellerReininghaus09}; we discuss the required modification for the study
of the spin-1 system in the appendix.

%%%%%%%%%%%%%%%%%%%%%%%%%%%%%%%%%%%%%%%%%%%
\section{Spin-1 Kondo dot}\label{sec:model}
%%%%%%%%%%%%%%%%%%%%%%%%%%%%%%%%%%%%%%%%%%%
\begin{figure}[t]
	\centering
	\includegraphics[width=0.4\textwidth]{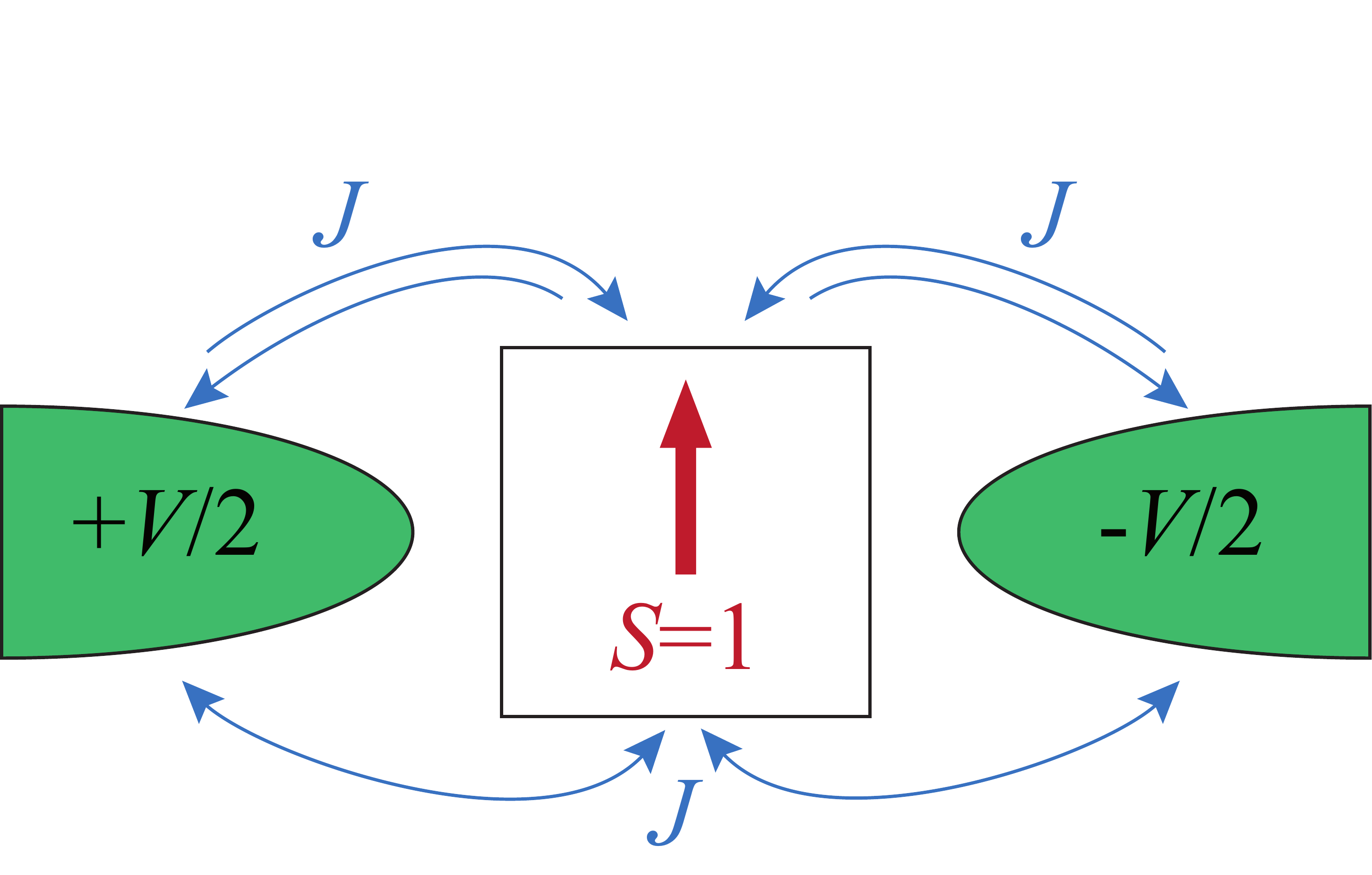}
	\caption{\label{fig:model}(Color online) Sketch of the spin-1 Kondo model 
	\eqref{eq:ham}.  The spin-1 on the quantum dot is coupled via exchange interactions 
	$J$ to the electron spins in the two reservoirs at different chemical potential 
	$\mu_{L/R}=\pm V/2$.}
\end{figure}
In this article we consider a spin-1 Kondo dot, i.e. a quantum dot whose internal degree
of freedom is given by a spin $S=1$, which is attached to two electronic reservoirs via 
a spin exchange interaction $J$. This set-up is relevant for recent transport experiments 
on $C_{60}$ molecules~\cite{Roch-08,Roch-09} and cobalt complexes~\cite{Parks-10}.
The spin-1 Kondo model can be regarded as a direct generalization 
of a spin-1/2 Kondo dot studied previously in Ref.~\onlinecite{SchoellerReininghaus09}, 
which itself is related to the single-orbital Anderson model by a Schrieffer--Wolff 
transformation.

Specifically, the Hamiltonian of the spin-1 Kondo dot is given by (see Fig.~\ref{fig:model})
\begin{equation}
  \begin{split}
		H=&\sum_{\alpha k\sigma}(\epsilon_k-\mu_\alpha)\,
		c_{\alpha k\sigma}^\dagger c_{\alpha k\sigma}+ h_0\,S^z \\
		&+\frac{J_0}{2\nu_0} \sum_{\alpha\alpha' k k' \sigma\sigma'}\vec{S}
		\cdot\vec{\sigma}_{\sigma\sigma'}
		c_{\alpha k\sigma}^\dagger c_{\alpha' k'\sigma'}.
  \end{split}
  \label{eq:ham}
\end{equation}
Here $c_{\alpha k\sigma}^\dagger$ and $c_{\alpha k\sigma}$ create and annihilate 
electrons with momentum $k$ and spin $\sigma=\uparrow,\downarrow$ in lead $\alpha=L,R$, 
and $\vec\sigma$ denotes the Pauli matrices. The chemical potentials in the leads are 
given by $\mu_{L/R}=\pm V/2$ thus applying a finite bias voltage $V=\mu_L-\mu_R$ to 
the dot. Furthermore we introduce an ultra-violet cutoff $D$ via the density of states 
\begin{equation}
	N(\omega)=\frac{\nu_0D^2}{D^2+\omega^2}
\end{equation}
and note that the exchange coupling $J_0$ is dimensionless in our convention.
The spin on the dot is subject to an
external magnetic field $h_0$ applied along the z-direction.
If we use its direction as the quantization axis, the
components of the spin-1 operator on the dot are explicitly given by
\begin{equation}
	\vec{S}=
	\left\lbrace
	\frac{1}{\sqrt{2}}
	\begin{pmatrix}
		0 & 1 & 0 \\
		1 & 0 & 1 \\
		0 & 1 & 0
	\end{pmatrix},
	\frac{1}{\sqrt{2}}
	\begin{pmatrix}
		0 & -i & 0 \\
		i & 0 & -i \\
		0 & i & 0
	\end{pmatrix},
	\begin{pmatrix}
		1 & 0 & 0 \\
		0 & 0 & 0 \\
		0 & 0 & -1
	\end{pmatrix}
	\right\rbrace.\label{eq:spin1-operator}
\end{equation}
The impurity spin on the dot is 
coupled via an antiferromagnetic exchange interaction $J_0>0$ to the electron spins 
in the two reservoirs. (We note that the situation of reservoir-dependent exchange couplings
$J_{\alpha\beta}$ follows as a straightforward generalization as long as the condition
$J_{LR}^2=J_{LL}J_{RR}$ is satisfied.)
We consider zero temperature and $h_0,V\ge0$.

%%%%%%%%%%%%%%%%%%%%%%%%%%%%%%%%%%%%%%%%%%%
\section{RTRG-FS analysis}\label{sec:RTRG}
%%%%%%%%%%%%%%%%%%%%%%%%%%%%%%%%%%%%%%%%%%%
As it is well known for Kondo models like \eqref{eq:ham} perturbation theory in the exchange 
coupling $J_0$ leads to logarithmic divergencies. In order to takle this problem
we use the RTRG-FS~\cite{Schoeller09} method, 
which is particularly suited to derive analytic results for the spin decay rates, 
the renormalized magnetic field, the dot magnetization as well as the differential conductance in non-equilibrium. Our
calculations follow the analysis of the spin-1/2 Kondo model performed in 
Ref.~\onlinecite{SchoellerReininghaus09}; we present here a brief summary to set up the
notations and discuss the essential modifications required
for the spin-1 system in the appendix.

We start with the time evolution of the density matrix of the full system $\rho(t)$, which is 
governed by the von Neumann equation
\begin{equation}
	\dot\rho(t)=-i \comm{H}{\rho(t)}.
	\label{eq:von_neumann_equation}
\end{equation}
This equation is formally solved by 
\begin{equation}
	\rho(t)=e^{-i H(t-t_0)}\rho(t_0)e^{i H(t-t_0)}=e^{-i L(t-t_0)}\rho(t_0),
\end{equation}
where the Liouvillian $L$ is defined via $L\,.=\comm{H}{\,.\,}$. We assume that for $t<t_0$ 
the reservoirs and quantum dot are decoupled and hence that the initial density matrix
at $t=t_0$ is of the form $\rho(t_0)=\rho_S(t_0)\rho_\textit{res}^L\rho_\textit{res}^R$. 
Here $\rho_\textit{res}^\alpha$ is the grand canonical density matrix of reservoir 
$\alpha$ including 
the chemical potential $\mu_\alpha$ and $\rho_S(t_0)$ denotes the initial density matrix 
of the dot system. The central object of our analysis is the reduced density matrix of the 
dot $\rho_S(t)$, which is obtained by tracing out the reservoir degrees of freedom. 
Instead of studying the time evolution of the density matrix directly we further perform a 
Laplace transform
\begin{eqnarray}
	\tilde{\rho}_S(z)
		&=&
			\int_{t_0}^\infty dt\,e^{i z(t-t_0)}\rho_S(t)\nonumber\\
		&=&
			\tr_\textit{res}\frac{i}{z-L(z)}\rho_S(t_0)\rho_\textit{res},
			\label{eq:reduced-densitymatrix}
\end{eqnarray}
where $z$ is the Laplace variable. In order to calculate 
$\tilde{\rho}_S(z)$ we expand in the exchange interaction, which yields a diagrammatic
representation in Liouville space. Resumming the resulting series one 
obtains~\cite{Schoeller09}
\begin{equation}
	\tilde\rho_S(z)=\frac{i}{z-L_S^\textit{eff}(z)}\rho_S(t_0),
\end{equation}
where the effective dot Liouvillian is given by
\begin{equation}
	L_S^\textit{eff}(z)=L_S^{(0)}+\Sigma(z).
	\label{eq:lseff}
\end{equation}
Here the first term $L_S^{(0)}\,.=h_0\comm{S^z}{\,.\,}$ describes the action of the applied 
magnetic field on the spin, whereas the second term $\Sigma(z)$ incorporates all interaction
effects of the coupling to the leads. From $L_S^\textit{eff}(z)$ we obtain the renormalized 
magnetic field, the decay rates,  and the reduced density matrix of the dot (and thus the 
dot magnetization) which are presented in Secs.~\ref{sec:magnetic-field}, \ref{sec:rates},
and~\ref{sec:magn}, respectively. A similar analysis can be performed for the current 
kernel yielding the current and differential conductance through the system (see 
Sec~\ref{sec:conductance}).

The RTRG-FS allows to calculate the effective dot Liouvillian and  
current kernel as well as other observables (see e.g. Ref.~\onlinecite{SS09})
in a controlled weak-coupling expansion in the renormalized exchange couplings $J$. 
For Kondo-type models this is done~\cite{Schoeller09,SchoellerReininghaus09,PS11} 
by first deriving 
and solving the poor-man's scaling (PMS) equations. In the second step one then expands
all quantities of interest in the effective coupling $J(\Lambda)$ while staying in the 
weak-coupling regime $J\ll 1$.

In a more physical picture, the RG procedure reduces the effective 
band width by integrating out the high-energy degrees of freedom. To accomplish this the 
initial band width $D$ is replaced by an effective band width $\Lambda$ which flows from 
$\Lambda_0\sim D$ to the physical scale $\Lambda_c$ which in turn is determined by the
restriction to the weak-coupling regime. The RG equations describing the flow of the 
interaction vertices and effective Liouvillian are obtained from the generic RG equations 
derived in Ref.~\onlinecite{SchoellerReininghaus09} by specifying to the spin-1 case
(see appendix for details). In particular, the renormalization of the exchange coupling in
leading order is governed by the PMS equation 
\begin{equation}
	\frac{dJ}{d\Lambda}=-\frac{2}{\Lambda}\,J^2,
	\label{eq:PMSeq}
\end{equation}
which is identical to the spin-1/2 model. The solution reads 
\begin{equation}
	J(\Lambda)=\frac{1}{2\ln{\frac{\Lambda}{T_K}}},\quad T_K=\Lambda_0 e^{-1/2 J_0},
	\label{eq:exchangecoupling}
\end{equation}
where the Kondo temperature is defined as the scale at which the effective coupling 
constant diverges. The scaling limit is obtained by taking $\Lambda_0\to\infty$ and 
$J_0\to 0$ while keeping $T_K$ constant. Below we will always perform the scaling limit
unless stated otherwise [e.g. in \eqref{eq:RMF1}].

As the PMS coupling diverges at $\Lambda=T_K$, the  weak-coupling condition 
$J(\Lambda)\ll 1$ can be translated into a condition for the physical cut-off scale 
\begin{equation}
	\Lambda_c=\max\lbrace\abs{z},V,h_0 \rbrace,
\end{equation}
which has to satisfy $\Lambda_c\gg T_K$. As $z\to 0$ for stationary quantities we 
thus recover the condition \eqref{eq:wcc}. In this regime a perturbative expansion of the 
Liouvillian and the current kernel in $J_c=J(\Lambda_c)$ is well-defined. 
Hereby the precise value of $\Lambda_c$
is irrelevant~\cite{SchoellerReininghaus09,SS09,Schoeller09} 
as changes $\Lambda_c\to\Lambda_c'$ with $\Lambda_c'/\Lambda_c\sim 1$ 
only lead to higher-order corrections in $J_c$ which will not be analyzed. 
The explicit calculation of  the Liouvillian and 
current kernel follows along the lines of
Ref.~\onlinecite{SchoellerReininghaus09} with all necessary modification required
for the spin-1 Kondo model discussed in App.~\ref{app:rg-calculation}. In the 
following we present the results for the renormalized magnetic field and g-factor, 
decay rates and time evolution, magnetization and anisotropy as well as the 
differential conductance through the system,
where we in particular focus on the differences to the spin-1/2 model.

%%%%%%%%%%%%%%%%%%%%%%%%%%%%%%%%%%%%%%%%%%%
\section{Results}\label{sec:results}
%%%%%%%%%%%%%%%%%%%%%%%%%%%%%%%%%%%%%%%%%%%
The systematic weak-coupling expansions for the effective dot Liouvillian 
$L_S^\textit{eff}(z)$ and the current kernel $\Sigma^I(z)$ involve terms of the order $J_c^3\,\ln$, 
where $\ln$ denotes terms 
of the form $\ln[i\Lambda_c/(z+\Delta+i\Gamma_i)]$ with some physical scale $\Delta=V,h_0,
V\pm h_0$ and $\Gamma_i\sim J_c^2\Delta$ the decay rates (see Sec.~\ref{sec:rates}).
From the dot Liouvillian one can derive the renormalized magnetic field as well as the decay
rates by determining the poles of the resolvent $1/[z-L_S^\textit{eff}(z)]$. If we use the spectral
decomposition $L_S^\textit{eff}(z)=\sum_i\,\lambda_i(z)\,P_i(z)$, where $\lambda_i(z)$ and 
$P_i(z)$ denote the eigenvalues and corresponding projectors respectively, 
they are determined by solving~\cite{Schoeller09,SchoellerReininghaus09}
\begin{equation}
z_i=\lambda_i(z_i)
\label{eq:selfconsistency-equation}
\end{equation}
where the real parts of $z_i$ yield oscillation frequencies in the time evolution of observables
and thus the renormalized magnetic field, while the imaginary parts lead to exponential
decay and thus determine the decay rates (see App.~\ref{app:liouvillian} for details).
Furthermore, the stationary reduced density matrix $\rho_S^\textit{st}$ is obtained by solving
\begin{equation}
L_S^\textit{eff}(i0^+)\,\rho_S^\textit{st}=0,
\label{eq:eqrhost}
\end{equation}
thus yielding the stationary dot magnetization and anisotropy (see Sec.~\ref{sec:magn}).
Finally the current through the system is derived from $\rho_S^\textit{st}$ and the current 
kernel.

%%%%%%%%%%%%%%%%%%%%%%%%%%%%%%%%%%%%%%%%%%%
\subsection{Renormalized magnetic field and g-factor}\label{sec:magnetic-field}
%%%%%%%%%%%%%%%%%%%%%%%%%%%%%%%%%%%%%%%%%%%
The renormalized magnetic field $h$, which emerges from 
the externally applied magnetic field $h_0$ due to screening processes from the electrons
in the leads, is determined by the real parts of the poles $z_i$ 
derived from the self-consistency equations \eqref{eq:selfconsistency-equation}.
We present explicit relations for the $z_i$'s in Eqs.~\eqref{eq:z1}--\eqref{eq:z89}. 
To first order in the exchange coupling we find
\begin{equation}
	h^{(1)}=\big[1-(J_c-J_0)\big]h_0,
	\label{eq:RMF1}
\end{equation}
which is identical~\cite{Garst-05,SchoellerReininghaus09,FritschKehrein10} 
to the renormalization of the magnetic field in the spin-1/2 Kondo model. Differences 
in the models manifest themselves in the logarithmic corrections to \eqref{eq:RMF1},
which yield
\begin{equation}
		h =h^{(1)} -\frac{J_c^2}{2}h\logTerm{-}{h}
			+\frac{J_c^2}{4}\big(V-h\big)\logTerm{-}{V-h}.
				\label{eq:RMF2}
\end{equation}
In the derivation of \eqref{eq:RMF2} we have neglected all terms in order $J_c^2$ that do
not contain logarithms at either $h=0$ or $h=V$, performed the scaling limit $J_0\to 0$, and 
further replaced $h_0\to h$ (which only leads to higher-order corrections).
In addition, we have introduced the short-hand notation 
\begin{equation}
	\mathcal{L}_-(x)=
		\ln\frac{\Lambda_c}{\sqrt{x^2+(\Gamma_t^1-\Gamma_q^2)^2}}+
		\ln\frac{\Lambda_c}{\sqrt{x^2+(\Gamma_q^1-\Gamma_q^2)^2}},
		\label{eq:log_minus}
\end{equation}
where the appearing decay rates $\Gamma_t^1$, $\Gamma_q^1$, and $\Gamma_q^2$ 
will be discussed in detail in the next section (see Fig.~\ref{fig:rates}). Here we already 
note that at resonance $V=h$ they satisfy $\Gamma_t^1\neq\Gamma_q^2$ and 
$\Gamma_q^1\neq\Gamma_q^2$, i.e. \eqref{eq:log_minus} is well-defined.

\begin{figure}[t]
	\centering
	\includegraphics[width=0.5\textwidth]{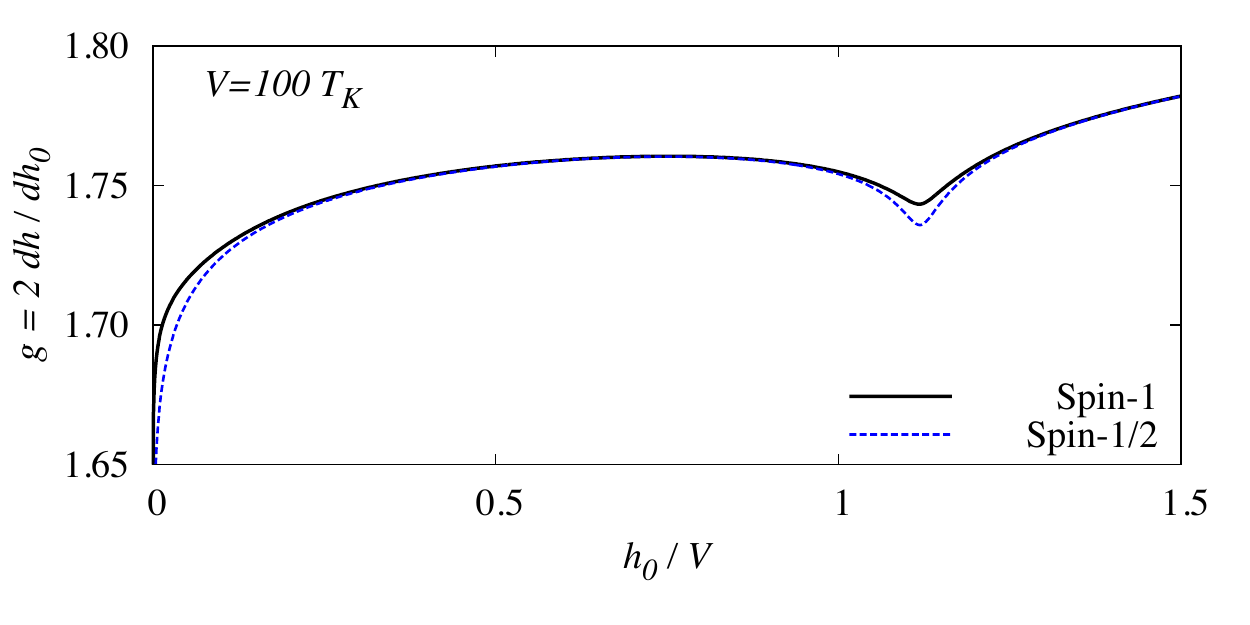}
	\caption{\label{fig:g-factor}(Color online) The g-factor $g=2dh/dh_0$ for the spin-1 
	(solid line) and the spin-1/2 (dashed line, see 
	Ref.~\onlinecite{SchoellerReininghaus09}) model for $V=100\,T_K$, 
	where $J_c=J(\Lambda_c)$ is given in Eq.~\eqref{eq:exchangecoupling} with 
	$\Lambda_c=\max\lbrace{V,h}\rbrace$. We observe 
	a logarithmic feature at $h=V$ which is less pronounced in the spin-1 model.}
\end{figure}

From the renormalized magnetic field \eqref{eq:RMF2} we can directly obtain the 
g-factor 
\begin{equation}
	g=2\frac{dh}{dh_0}=2(1-J_c)-J_c^2\mathcal{L}_-(h)-\frac{J_c^2}{2}\mathcal{L}_-(V-h),
\end{equation}
which is plotted in Fig.~\ref{fig:g-factor}. Compared to its bare value the 
g-factor is reduced due to the screening of the spin on the dot by the electron spins in the leads. We further observe two logarithmic features at $h=0$ 
and $h=V$. Due to the reduction of the renormalized magnetic field compared to $h_0$ the 
latter appears at $h_0>V$, however, up to logarithmic corrections in order $J_c^2$ the
position equals the spin-1/2 situation. Differences  show up in the line shapes close to $h=0$ and $h=V$,
where the spin-1 result clearly shows a larger broadening. The reason is that in the 
spin-1 case both broadenings $\Gamma_q^1$ and $\Gamma_t^1$ (which satisfy 
$\Gamma_q^1>\Gamma_t^1$), appear in the logarithms, while in the spin-1/2 case one 
finds~\cite{SchoellerReininghaus09} $\Gamma_t^1=\Gamma_q^1$. Away from these positions,
however, the g-factors are identical.

%%%%%%%%%%%%%%%%%%%%%%%%%%%%%%%%%%%%%%%%%%%
\subsection{Decay rates and time evolution}\label{sec:rates}
%%%%%%%%%%%%%%%%%%%%%%%%%%%%%%%%%%%%%%%%%%%
In this section we investigate the imaginary parts of the poles $z_i$ obtained by solving 
\eqref{eq:selfconsistency-equation}. As we will show they lead to an exponential decay 
in the time evolution of observables and thus constitute the decay rates of the system.
As the local Hilbert space as well as the corresponding Liouville space of the spin-1 dot 
has a larger dimension than its spin-1/2 analog, we expect the appearance of additional
decay rates in the spin-1 case. In the following we first analyze these rates in the case of 
vanishing magnetic field and then consider the case of a finite field.

%%%%%%%%%%%%%%%%%%%%%%%%%%%%%%%%%%%%%%%%%%%
\subsubsection{No magnetic field $h_0=0$}
%%%%%%%%%%%%%%%%%%%%%%%%%%%%%%%%%%%%%%%%%%%
For vanishing magnetic field $h_0=0$ the model is fully spin isotropic. Thus the local 
Liouville space of the dot can be decomposed into irreducible representations of SU(2).
For a spin-1/2 system these are a singlet and a triplet 
(according to $\tfrac{1}{2}\otimes\tfrac{1}{2}=0\oplus 1$), while for the spin-1 system
one finds a singlet, triplet, and quintet ($1\otimes 1=0\oplus 1\oplus 2$). Therefore the
effective dot Liouvillian of the spin-1 model \eqref{eq:ham} decomposes as 
\begin{equation}
	L_S^\textit{eff}(z)=-i f_{s}(z)L^{s}-i f_{t}(z)L^{t}-i f_{q}(z)L^{q},
	\label{eq:Leffh0}
\end{equation}
where the superoperators $L^{s}$, $L^{t}$, and $L^{q}$ project onto the respective
subspaces (see App.~\ref{app:liouvillian}). In leading order the coefficients in 
\eqref{eq:Leffh0}
read $f_s(z)=0$, $f_t(z)=\pi J_c^2V$, and $f_q(z)=3\pi J_c^2V$. As there is no dependence
on the Laplace variable $z$ the self-consistency equations 
\eqref{eq:selfconsistency-equation} are trivially solved by
$z_i=-i\Gamma_i$ ($i=s,t,q$) with
\begin{equation}
	\Gamma_s =0,\quad\Gamma_{t}=\pi J_c^2V,\quad\Gamma_{q}=3\pi J_c^2V.
	\label{eq:gammah0}
\end{equation}
Obviously the corresponding subspaces in Liouville space are 
1-dimensional ($\Gamma_s$), 3-dimensional ($\Gamma_t$), and 5-dimensional
($\Gamma_q$). The difference to the spin-1/2 model~\cite{SchoellerReininghaus09} 
is the appearance of the additional rate $\Gamma_q$. 
Similarly the decay rates in a general spin-$S$ Kondo dot in the absence of a magnetic field
are given by
\begin{equation}
	\Gamma_j=\frac\pi2j(j+1)J_c^2V\quad\text{for }j=0,1,\ldots,2S.
\end{equation}

In order to interpret the result \eqref{eq:gammah0} we investigate the time evolution of the 
reduced density matrix, which is given by the inverse Laplace transform 
of \eqref{eq:reduced-densitymatrix}
\begin{equation}
  \rho_D(t)=\frac{i}{2\pi}\int\nolimits_{-\infty+i0^+}^{\infty+i0^+}
  \frac{dz\,e^{-izt}}{z-L_D^\textit{eff}(z)}\,\rho_D(0).
  \label{eq:inLT}
\end{equation}
The initial density matrix is assumed to be of the form
$\rho(t_0=0)=\rho_S(0)\rho_\textit{res}^L\rho_\textit{res}^R$ with 
\begin{equation}
  \rho_S(0)=\left(\begin{array}{ccc}\rho_{11}&\rho_{10}&\rho_{1-1}\\
    \rho_{01}&\rho_{00}&\rho_{0-1}\\\rho_{-11}&\rho_{-10}&\rho_{-1-1}
  \end{array}\right).
  \label{eq:rhoD0}
\end{equation}
We perform the integral in \eqref{eq:inLT} by using the decomposition \eqref{eq:Leffh0} 
for the Liouvillian. In leading order this yields 
\begin{equation}
	\rho_S(t)=\bigl[L^{s}+e^{-\Gamma_tt}\,L^{t}+e^{-\Gamma_qt}\,L^{q}\bigr]\rho_S(0).
	\label{eq:rhot}
\end{equation}
We note that the time evolution is governed by purely exponential decays. This is due to the 
absence of a $z$-dependence of $L_S^\textit{eff}$ in leading order; in general the branch 
cuts of $L_S^\textit{eff}$ due to logarithmic terms will lead to additional power-law 
corrections~\cite{PSS10}. The time evolution of the spin operator is now readily obtained
from \eqref{eq:rhot}
\begin{eqnarray}
	\average{\vec{S}(t)}&\propto&e^{-{\Gamma}_{t}t}\text{, e.g.}\nonumber\\
	\average{S^z(t)}&=&\bigl(\rho_{11}-\rho_{-1-1}\bigr)e^{-{\Gamma}_{t}t}.\label{eq:averageSZ}
\end{eqnarray}
Thus the magnetization decays only with the triplet rate $\Gamma_t$, which is
identical~\cite{PSS10} to the spin-1/2 model. However, in the spin-1 model one can 
construct a second non-trivial operator on the dot, namely the anisotropy 
operator~\cite{VarshalovichMoskalevKhersonskii88} (or spherical quadrupole tensor) 
with components
\begin{eqnarray}
	T_{2\pm2}&=&\frac12[S^xS^x-S^yS^y\pm i(S^xS^y+S^yS^x)],\\
	T_{2\pm1}&=&\mp\frac12[S^xS^z+S^zS^x\pm i(S^yS^z+S^zS^y)],\\
	T_{20}&=&\sqrt{\frac32}S^zS^z-\sqrt{\frac23}\mathbb{1}.
\end{eqnarray}
($T$ is related~\cite{VarshalovichMoskalevKhersonskii88} to the traceless quadrupole
tensor $Q_{ij}$ introduced in Refs.~\onlinecite{SothmannKoenig10,Baumgaertel-11}
to study the transport through spin valves.)
Physically $T_{20}$ measures the anisotropy along the z-direction. 
$\average{T_{20}} > 0$ shows a tendency to align the impurity-spin along the z-axis, 
while $\average{T_{20}} < 0$ indicates an alignment in the xy-plane. We further note that
in the spin-1/2 case one finds $T_{20}\propto\mathbb{1}$, i.e. the anisotropy operator is trivial.
From \eqref{eq:rhot} we can easily infer the time evolution of $T$
\begin{eqnarray}
  \average{T(t)}
		&\propto&e^{-{\Gamma}_{q}t}\text{, e.g.}\nonumber\\
	\average{T_{20}(t)}
		&=&
		\frac1{\sqrt{6}}(\rho_{11}-2\rho_{00}+\rho_{-1-1})e^{-\Gamma_{q}t}.
		\label{eq:averageT20}
\end{eqnarray}
Thus the situation in the absence of a magnetic field is quite simple: The magnetization 
decays with the triplet rate $\Gamma_t$, whereas
the quintet rate $\Gamma_q$ governs the time evolution of the anisotropy 
operator. In the next section we consider how this changes if the spin rotational invariance
is broken by the application of a finite magnetic field.

%%%%%%%%%%%%%%%%%%%%%%%%%%%%%%%%%%%%%%%%%%%
\subsubsection{Finite magnetic field $h_0>0$}
%%%%%%%%%%%%%%%%%%%%%%%%%%%%%%%%%%%%%%%%%%%
\begin{figure}[t]
	\centering
	\includegraphics[width=0.5\textwidth]{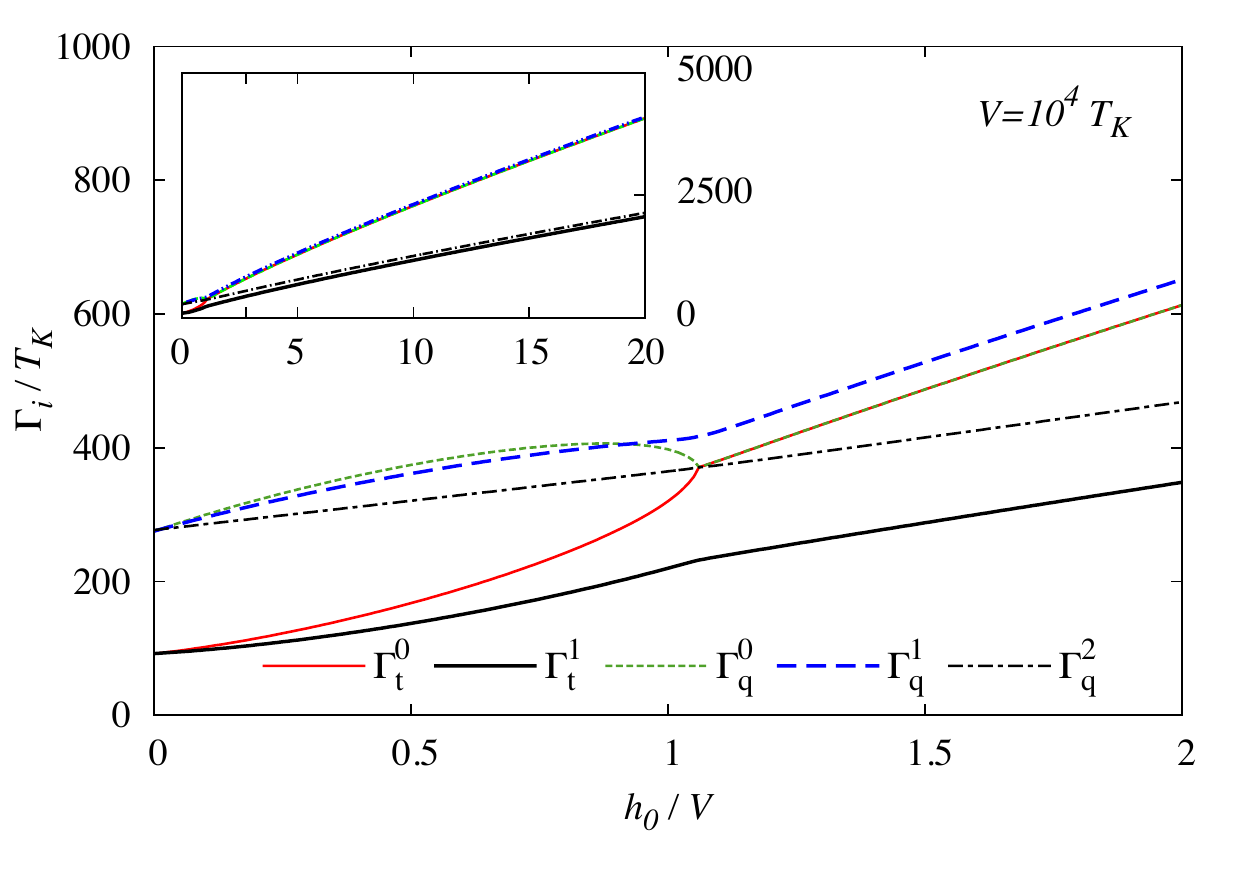}
	\caption{\label{fig:rates}(Color online) The decay rates \eqref{eq:rates} 
	for $V=10^4\,T_K$, 
	where $J_c=J(\Lambda_c)$ is given in Eq.~\eqref{eq:exchangecoupling} with 
	$\Lambda_c=\max\lbrace{V,h}\rbrace$. 
	We observe the splitting of the triplet and quintet rate into five 
	rates, of which two ($\Gamma_t^0$ and $\Gamma_q^0$) merge for $h\ge V$. Inset: For 
	$h_0\gg V$ there are two rates governing the decay along and perpendicular 
	to the z-direction.}
\end{figure}
The breaking of the spin rotational invariance results in a decomposition of the triplet and 
quintet subspaces in the local Liouville space of the dot. Hence the dot Liouvillian does 
no longer take the simple form \eqref{eq:Leffh0} and the system possesses more than two
decay rates. Specifically we find that the rates \eqref{eq:gammah0} split up as
\begin{equation}
\Gamma_t\to\Gamma_t^0,\Gamma_t^1,\quad
\Gamma_q\to\Gamma_q^0,\Gamma_q^1,\Gamma_q^2.\label{eq:rates}
\end{equation}
Compared to the spin-1/2 model, where at finite magnetic field two different rates 
(the longitudinal and transverse spin relaxation rates usually denoted by $\Gamma_1$
and $\Gamma_2$ respectively) exist, we find five decay rates in the spin-1 system.
As each rate corresponds to a relaxation or decoherence time, 
the time evolution towards the stationary state will be more complex in the spin-1 model 
(see below). 

In the small-field limit $h_0\ll V$ the splitting of the rates is explicitly given by
\begin{equation}
\Gamma_t^0=\pi J_c^2V+\pi J_c^2h,\quad
\Gamma_t^1=\pi J_c^2V+\frac{\pi}{2} J_c^2h,
\label{eq:ratest}
\end{equation}
and 
\begin{equation}
\begin{split}
&\Gamma_q^0=3\pi J_c^2V+3\pi J_c^2h,\quad
\Gamma_q^1=3\pi J_c^2V+\frac{5}{2}\pi J_c^2h,\\
&\Gamma_q^2=3\pi J_c^2V+\pi J_c^2h.
\end{split}
\label{eq:ratesq}
\end{equation}
We note that the magnetic-field dependence of the triplet rates \eqref{eq:ratest} is 
identical to the spin-1/2 model~\cite{SchoellerReininghaus09} with the identifications 
$\Gamma_t^0\to\Gamma_1$ and $\Gamma_t^1\to\Gamma_2$.
For arbitrary magnetic fields the rates have been obtained by numerically solving the
self-consistency equations \eqref{eq:gammaTQ0}--\eqref{eq:gammaQ2}, the result is 
shown in Fig.~\ref{fig:rates}. We observe in particular that two of the rates 
($\Gamma_t^0$ and $\Gamma_q^0$) merge at $h=V$. 
We note that \eqref{eq:log_minus} is well-defined at resonance
as $\Gamma_t^1\neq\Gamma_q^2$ and $\Gamma_q^1\neq\Gamma_q^2$.

The time evolution of the reduced density matrix follows from \eqref{eq:inLT} using the 
spectral decomposition  $L_S^\textit{eff}(z)=\sum_i\lambda_i(z)P_i(z)$ of  the dot Liouvilian. 
If we further approximate this by expanding around the poles of the resolvent, 
$L_S^\textit{eff}(z)\approx\sum_i\lambda_i(z_i)P_i(z_i)$ and 
use Eq.~\eqref{eq:selfconsistency-equation}, we obtain
\begin{equation}
\rho_S(t)=\sum_ie^{-iz_it}\,P_i(z_i)\,\rho_S(t=0).
\label{eq:rht}
\end{equation}
Here the real parts of the $z_i$ yield oscillations with the oscillation frequency determined 
by the renormalized magnetic field, while the imaginary parts correspond to the decay rates
\eqref{eq:rates} [see Eqs.~\eqref{eq:z1}--\eqref{eq:z89}]. The first pole $z_1=0$ yields a 
stationary contribution, i.e. $P_1(0)$ is the projector onto the stationary density matrix
$\rho_S^\textit{st}$. From \eqref{eq:rht} we obtain for example
\begin{equation}
	\average{T_{2\pm2}(t)}=e^{\pm2ih t}e^{-\Gamma_q^2 t}\rho_{\mp1\pm1}.
\end{equation}
Corrections to \eqref{eq:rht} can be calculated along the lines of 
Refs.~\onlinecite{PSS10,KAPSBMS10,APSSB11}.
We expect additional terms oscillating with the freqencies $V$, $h\pm V$, and $2h\pm V$
as well as power-law corrections to the exponential decays.

A better understanding of the time evolution and the relevant decay rates can be obtained
by studying the large-field limit, $h_0\gg V$. In this regime we find only two effective 
rates which are related to the five original ones by (see inset in Fig.~\ref{fig:rates})
\begin{eqnarray}
	\Gamma_t^1, \Gamma_q^2 &\rightarrow& \Gamma_\perp=2\pi J_c^2h,\label{eq:rateperplargeh}\\
	\Gamma_t^0, \Gamma_q^0, \Gamma_q^1 &\rightarrow& \Gamma_z=4\pi J_c^2h.
	\label{eq:ratezlargeh}
\end{eqnarray}
[We note that although $\Gamma_t^1\to\Gamma_q^2$ the logarithmic terms
in \eqref{eq:RMF2} are well defined as we are off resonance.]
The time evolution of the magnetization is now given by
\begin{eqnarray}
\langle S^z(t)\rangle&=&-\bigl(1-e^{-\Gamma_z t}\bigr)+
e^{-\Gamma_z t}\bigl(\rho_{11}-\rho_{-1-1}\bigr),\\
\langle S^\pm(t)\rangle&=&\frac{e^{\pm ih t}}{\sqrt{2}}
\bigl(e^{-\Gamma_\perp t}+e^{-\Gamma_z t}\bigr)
\bigl(\rho_{\mp 10}+\rho_{0\pm 1}\bigr),\quad
\end{eqnarray}
with $S^\pm=S^x\pm iS^y$ and the initial density matrix of the dot given by \eqref{eq:rhoD0},
while the anisotropy operator relaxes as 
\begin{eqnarray}
\langle T_{20}(t)\rangle&=&\frac{1}{\sqrt{6}}\bigl(1-e^{-\Gamma_z t}\bigr)\nonumber\\*
& &+\frac{e^{-\Gamma_z t}}{\sqrt{6}}\bigl(\rho_{11}-2\rho_{00}+\rho_{-1-1}\bigr),\\
\langle T_{2\pm 1}(t)\rangle&=&\frac{e^{\pm ih t}}{2\sqrt{2}}
	\bigl(e^{-\Gamma_\perp t}+e^{-\Gamma_z t}\bigr)
	\bigl(\rho_{\mp 10}-\rho_{0\pm 1}\bigr),\quad\\
\langle T_{2\pm 2}(t)\rangle&=&e^{\pm2ih t}e^{-\Gamma_\perp t}\rho_{\mp1\pm1}.
\end{eqnarray} 
We observe that the decay of the expectation values of the diagonal operators 
$S^z$ and $T_{20}$ is purely governed by the rate $\Gamma_z$, while 
$\langle T_{2\pm 2}(t)\rangle$ decays purely with $\Gamma_\perp$. Thus we conclude 
that $\Gamma_z$ corresponds to the longitudinal spin relaxation rate of the spin-1/2 
system (usually denoted by $\Gamma_1$), while $\Gamma_\perp$ corresponds to the 
transverse one (usually denoted by $\Gamma_2$). Comparing the explicit values \eqref{eq:rateperplargeh} and
\eqref{eq:ratezlargeh} with the spin-1/2 result~\cite{SchoellerReininghaus09} shows that 
both rates $\Gamma_\perp$ and $\Gamma_z$ are twice as large as their spin-1/2 
counterparts (i.e. their ratio is identical in the spin-1/2 and spin-1 model).
We stress, however, that the time evolution in the spin-1 model is more complex, e.g.
in the decay of $\langle S^\pm(t)\rangle$ and $\langle T_{2\pm 1}(t)\rangle$
both rates $\Gamma_z$ and $\Gamma_\perp$ appear.

%%%%%%%%%%%%%%%%%%%%%%%%%%%%%%%%%%%%%%%%%%%
\subsection{Magnetization and Anisotropy}\label{sec:magn}
%%%%%%%%%%%%%%%%%%%%%%%%%%%%%%%%%%%%%%%%%%%
In this section we analyze the stationary values of magnetization and anisotropy before 
turning to the differential conductance in the next section. The stationary density matrix 
of the dot is
obtained by solving \eqref{eq:eqrhost}. Introducing the observables magnetization and
anisotropy by
\begin{equation}
M=\average{S^z}_{st},\quad A=\average{T_{20}}_{st},
\end{equation}
the stationary density matrix can be written as 
\begin{equation}
	\rho_S^\textit{st}=\frac13\mathbb{1}+\frac{M}2S^z+A\,T_{20}.
	\label{eq:stationary-density-matrix}
\end{equation}
The knowledge of the effective dot Liouvillian enables us to derive analytic results for 
the magnetization and anisotropy including the leading logarithmic corrections, which 
are given by 
\begin{equation}
	M=\frac{4 f_1 f_2}{(f_1)^2+3(f_2)^2},\quad
	A=\frac1{\sqrt6}\frac{4(f_1)^2}{(f_1)^2+3(f_2)^2},
	\label{eq:MA}
\end{equation}
with
\begin{eqnarray*}
	f_1
		&=&
			2\pi J_c^2 h + 2\pi J_c^3 h\logTerm{1}{h}-\pi J_c^3(V-h)\logTerm{1}{V-h}\\
	f_2
		&=&
			-\frac\pi4 J_c^2\Bigl(2V+6h+\abs{V-h}_1\Bigr)\\
			&&\!\!\!\!\!\!\!\!\!\!-\pi J_c^3h\logTerm{1}{h}-\frac{\pi}{4}J_c^3
			\Big[\abs{V-h}_1-2(V-h)\Big]\logTerm{1}{V-h}
\end{eqnarray*}
as well as 
\begin{eqnarray}
	\logTerm{1}{x}&=&
		\ln\frac{\Lambda_c}{\sqrt{x^2+(\Gamma_{t\!\!\!\phantom{q}}^1)^2}}+
		\ln\frac{\Lambda_c}{\sqrt{x^2+(\Gamma_q^1)^2}}
		\label{eq:log}\\
	\abs{x}_1&=&
	\frac2\pi x \left(\arctan\frac{x}{\Gamma_{t\!\!\!\phantom{q}}^1}+
	\arctan\frac{x}{\Gamma_{q}^1}\right).\label{eq:abs1}	
\end{eqnarray}
In the derivation of \eqref{eq:MA} we have neglected all terms in order $J_c^3$ that do
not contain logarithms at either $h=0$, $V=0$, or $h=V$.
We note that the decay rates $\Gamma_t^1$ and $\Gamma_q^1$ cut-off the 
logarithmic divergencies present in bare perturbation theory. As we will see below these
two rates also appear in the logarithmic corrections to the differential conductance.\\

\begin{figure}[t]
	\centering
	\includegraphics[width=0.5\textwidth]{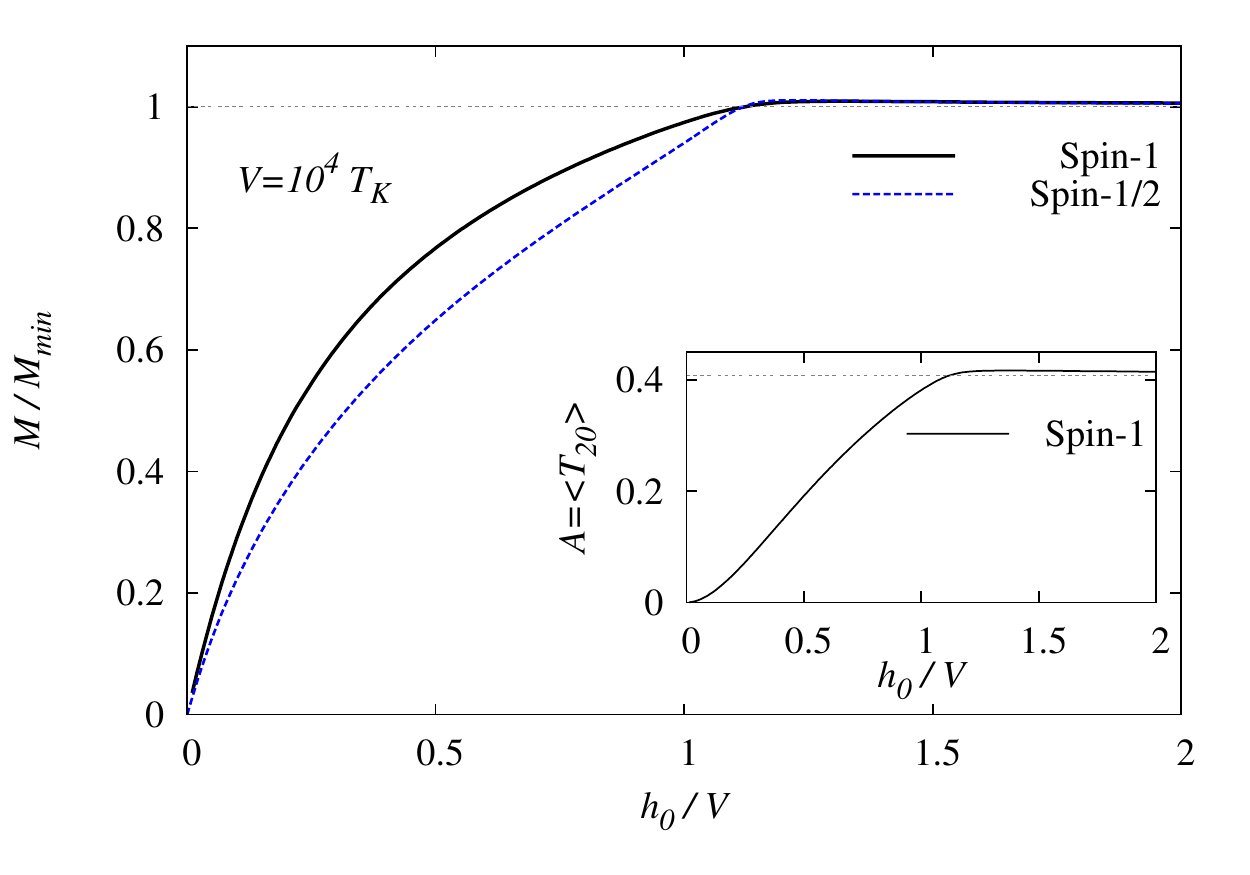}
	\caption{\label{fig:magnetization}(Color online) The relative magnetization 
	$M/M_\text{min}$ for the spin-1 (solid line) and the spin-1/2 
	(dashed line)~\cite{SchoellerReininghaus09} model for $V=10^4\,T_K$, 
	where $J_c=J(\Lambda_c)$ is given in Eq.~\eqref{eq:exchangecoupling} with 
	$\Lambda_c=\max\lbrace{V,h}\rbrace$.
	Inset: Anisotropy $A$ for the spin-1 model.}
\end{figure}

Specifically we find that for large magnetic fields $h>V$ the spin is completely 
aligned along the field, i.e. $M=-1$ and $A=1/\sqrt{6}$. Close to the resonance,
i.e. at $h<V$, $V-h\ll h$ we find by expanding \eqref{eq:MA} up to 
$O(J\ln)$
\begin{eqnarray}
	M
		&=&
			-1+\frac14\frac{V-h}{h}+\frac{J_c}{4}\frac{V-h}{h}\logTerm{1}{V-h},\\
	A &=&
			\frac1{\sqrt{6}}\left[
				1-\frac34\frac{V-h}{h}-\frac{3J_c}{4}\frac{V-h}{h}\logTerm{1}{V-h}
			\right].
\end{eqnarray}
Similarly in the limit of small magnetic fields, $h\ll V$, we obtain
\begin{eqnarray}
	M
		&=&
			-\frac{8h}{3V}\bigl[1+J_c\,\logTerm{1}{h}\bigr],\\
	A &=&\frac{16}{3\sqrt{6}}\frac{h^2}{V^2}\bigl[1+2J_c\,\logTerm{1}{h}\bigr].
\end{eqnarray}
The magnetization (together with the spin-1/2 result) and anisotropy are shown in 
Fig.~\ref{fig:magnetization}.

\begin{figure}[t]
	\centering
	\includegraphics[width=0.5\textwidth]{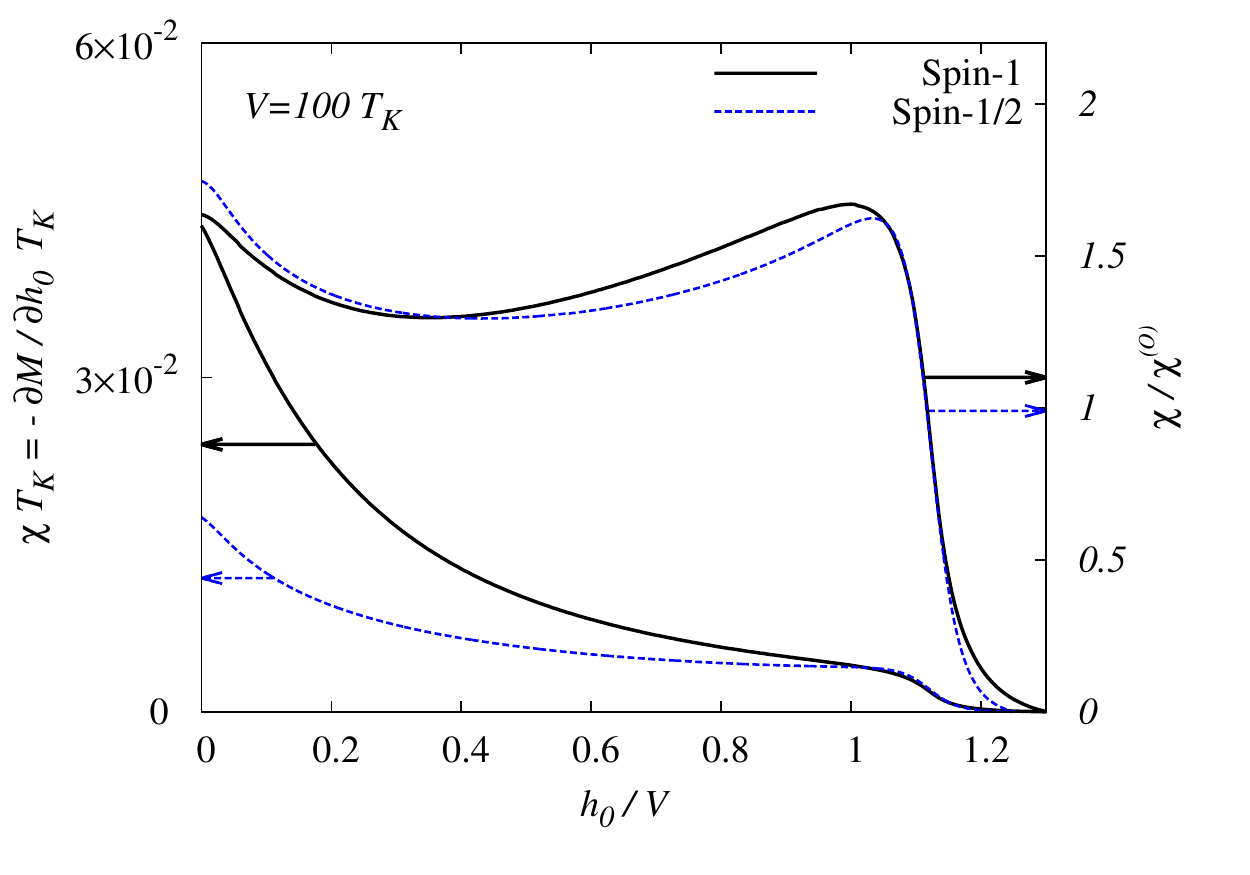}
	\caption{\label{fig:sus_limit}(Color online) Left axis: The susceptibility 
	$\chi=-\frac{\partial M}{\partial h_0}$ for the spin-1 (solid lines) and 
	spin-1/2 (dashed lines)~\cite{SchoellerReininghaus09} models for $V=100\,T_K$, 
	where $J_c=J(\Lambda_c)$ is given in Eq.~\eqref{eq:exchangecoupling} with 
	$\Lambda_c=\max\lbrace{V,h}\rbrace$. 
	Right axis: $\chi/\chi^{(0)}$, where $\chi^{(0)}$ is the result in 
	$O(J_c^0)$, i.e. without logarithmic corrections. We note that the 
	logarithmic corrections at $h=0$ and $h=V$ are less pronounced in the spin-1 model.}
\end{figure}

The corresponding magnetic susceptibility is readily obtained from $M$. At resonance 
and for small fields we find
\begin{eqnarray}
	\chi_{h=V}
		&=&-\frac{\partial M}{\partial h_0}\bigg|_{h\to V^-}
			=\frac{1}{4V}\bigl[1+J_c\,\logTerm{1}{V-h}\bigr],
			\label{eq:sus_h_voltage}\\
	\chi_{h=0}
		&=&-\frac{\partial M}{\partial h_0}\bigg|_{h\to 0}
		=\frac{8}{3V}\bigl[1+J_c\,\logTerm{1}{h}\bigr].\label{eq:sus_small_h}
\end{eqnarray}
Except for the logarithmic broadening the susceptibility at $h=V$ is identical to the spin-1/2 
case~\cite{SchoellerReininghaus09}, whereas in the limit $h\to 0$ they differ by a relative
factor $8/3$. We note that the same factor also appears in the susceptibility of an isolated 
spin $S$ at finite temperature. 
In Fig.~\ref{fig:sus_limit} we plot the susceptibility for the spin-1 model as well as the spin-1/2
system~\cite{SchoellerReininghaus09}. In order to enhance the visibility of the
logarithmic terms we further plot $\chi/\chi^{(0)}$, where $\chi^{(0)}$ is the result in 
$O(J_c^0)$. In comparison to the spin-1/2 model the logarithmic corrections at $h\to 0$ and 
$h\approx V$ are broadened as they contain two decay rates [see \eqref{eq:log}] of which 
one is larger than the rate appearing in the spin-1/2 susceptibility. We expect this to
be a generic feature of the spin-$S$ Kondo dot, i.e. the logarithmic corrections will be 
most pronounced in the $S=1/2$ model and become more and more broadened in the 
large-$S$ limit.

%%%%%%%%%%%%%%%%%%%%%%%%%%%%%%%%%%%%%%%%%%%
\subsection{Differential Conductance}\label{sec:conductance}
\begin{figure}[t]
	\centering
	\includegraphics[width=0.5\textwidth]{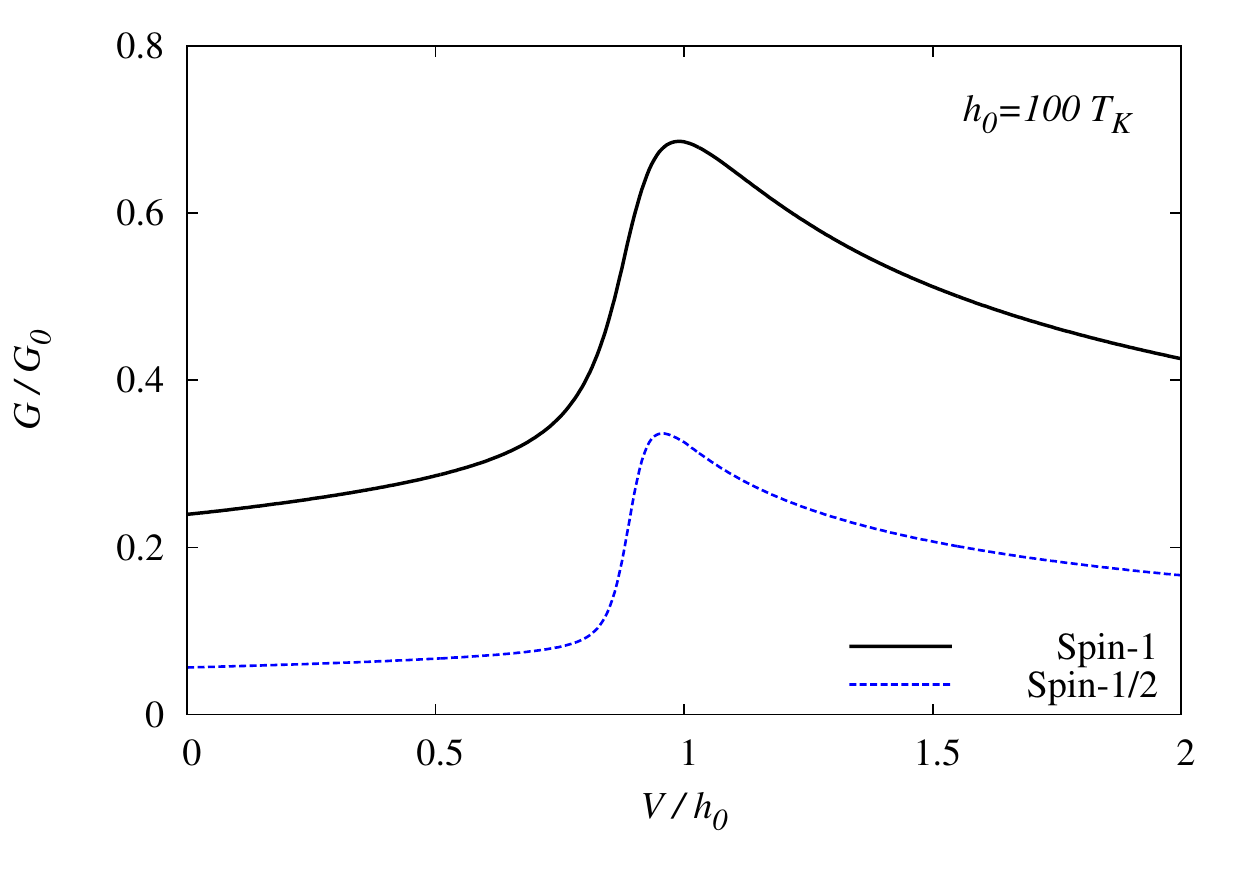}
	\caption{\label{fig:conductance}(Color online) The differential conductance 
	$G=\frac{dI}{dV}$ in the spin-1 (solid line) and spin-1/2
	(dashed line)~\cite{SchoellerReininghaus09} model
	 for $h_0=100\,T_K$, 
	where $J_c=J(\Lambda_c)$ is given in Eq.~\eqref{eq:exchangecoupling} with 
	$\Lambda_c=\max\lbrace{V,h}\rbrace$. 
	We observe that the linear conductance 
	$G(V\to 0)$ in the spin-1 model takes four times the value of the one in 
	the spin-1/2 model
	[see Eq.~\eqref{eq:linearG}] and that the onset of inelastic cotunneling processes 
	at $V=h$ is sharper in the spin-1/2 model.}
\end{figure}

%%%%%%%%%%%%%%%%%%%%%%%%%%%%%%%%%%%%%%%%%%%
Finally, we discuss the current and conductance through the system. We define the 
current as the change of the electron number in the left reservoir, 
$I^L=-\frac{d}{dt}N_\textit{res}^L$. Because the number of electrons on the dot is fixed, 
the current in the right reservoir is accordingly given by $I^R=-I^L$. Thus we will omit 
the superscript in the following, $I\equiv I^L$. Within the RTRG-FS formalism the 
stationary current is given by~\cite{Schoeller09}
\begin{equation}
	I\equiv\average{I}^\textit{st}=-i\tr_S\Sigma^I(i0^+)\rho_S^\textit{st},
	\label{eq:stationary_current}
\end{equation}
where the current kernel $\Sigma^I(z)$ can be derived in complete analogy to the
effective dot Liouvillian.  
We calculate $\Sigma^I$ up to order $J_c^3\ln$, thus together with the knowlegde 
of $\rho_S^\textit{st}$ up to this order we are able to derive analytic results for current and
conductance including the leading logarithmic corrections. The current is explicitly given by
\begin{eqnarray}
	I
		&=&
			f_1^I+M\,f_M^{I}+\frac{1}{\sqrt{6}}A\,f_A^{I}.
			\label{eq:current_explicit}
\end{eqnarray}
where $M$ and $A$ are given by \eqref{eq:MA} and
\begin{eqnarray}
	f_1^{I}
		&=&2\pi J_c^2 V\\
		&&
			\!\!\!\!\!\!+\frac43\pi J_c^3\Bigl[(V-h)\logTerm{1}{V-h}+V\logTerm{0}{V}\Bigr],\nonumber\\
	f_M^{I}
		&=&
			\frac\pi4J_c^2\bigl(2V+2h-\abs{V-h}_1\bigr)\\
		&&
			\!\!\!\!\!\!+\frac\pi4 J_c^3
			\Bigl[
				4V\logTerm{0}{V}+ 4h\logTerm{1}{h}
				-\abs{V-h}_1\logTerm{1}{V-h}
			\Bigr],\nonumber\\
	f_A^{I}
		&=&
			\pi J_c^3\Bigl[(V-h)\logTerm{1}{V-h}-2V\logTerm{0}{V}\Bigr],
\end{eqnarray}
with the logarithmic terms given by Eq.~\eqref{eq:log} and 
\begin{equation}
	\logTerm{0}{x}=
		\ln\frac{\Lambda_c}{\sqrt{x^2+(\Gamma_{t\!\!\!\phantom{q}}^0)^2}}+
		\ln\frac{\Lambda_c}{\sqrt{x^2+(\Gamma_q^0)^2}}.
		\label{eq:log0}	
\end{equation}
Fig.~\ref{fig:conductance} shows
the differential conductance as a function of the applied bias voltage $V$ derived from Eq.~\eqref{eq:current_explicit}.

We find for small bias voltage $V\ll h$ that there are no logarithmic corrections 
and the current is governed solely by elastic cotunneling processes,
\begin{equation}
	I=\pi J_c^2\,V,\quad G=\frac{dI}{dV}=\pi J_c^2.
	\label{eq:linearG}
\end{equation}
This is four times the result in the spin-1/2 model due to the additionally available transport 
channels. (In the spin-S dot the elastic cotunneling current will be given by 
$I=\tfrac{\pi}{6}(2S+1)S(S+1)J_c^2V$.)

On the other hand, close to the resonance $V\approx h$ we observe a jump in the 
differential conductance which is due to the onset of inelastic cotunneling processes
(see also Refs.~\onlinecite{Paaske-06,RouraBasAligia09,RouraBasAligia10,Florens-11}).
Explicitly we find
\begin{equation}
	\frac{G}{G_0}=
	\begin{cases}
		2\pi^2J_c^2+2\pi^2J_c^3\logTerm{1}{V-h}&\text{for }V<h,\\[2mm]
		\frac92\pi^2J_c^2+\frac{9}{2}\pi^2J_c^3\logTerm{1}{V-h}&\text{for }V>h,
	\end{cases}
	\label{eq:conductance}
\end{equation}
where we assumed $|V-h|\ll h$ in both cases, and $G_0=1/2\pi$ is the 
conductance quantum. Furthermore, from \eqref{eq:conductance}
we deduce the height of the jump at $V=h$
\begin{equation}
	\frac{\Delta G}{G_0}=\frac52\pi^2 J_c^2\bigl[1+J_c\,\logTerm{1}{V-h}\bigr].
\end{equation}
We note that the analog prefactor in the spin-1/2 model 
equals~\cite{SchoellerReininghaus09} $3/2$. The physical origin of this jump is the onset
of inelastic cotunneling processes: For finite magnetic field the spin states on the dot are 
separated by the energy $h$. Thus for $V<h$ only elastic cotunneling processes are possible
in which electrons are transferred from the left to the right lead via virtual states on the dot 
without flipping its spin state. Hence the initial and final state of the dot have the same energy.
For $V>h$, however, also inelastic cotunneling processes will contribute in which the spin 
on the dot is flipped (thus requiring the energy $h$). In analogy to the
susceptibility we observe that the logarithmic corrections are less pronounced in the 
spin-1 model.

Finally we would like to add that the current noise in the spin-1/2 and spin-1 models are
qualitatively similar, i.e. differences appear only in the decay rates broadening the 
logarithmic corrections~\cite{Mueller}.

%%%%%%%%%%%%%%%%%%%%%%%%%%%%%%%%%%%%%%%%%%%
\section{Anisotropic model}\label{sec:anisotropic-model}
%%%%%%%%%%%%%%%%%%%%%%%%%%%%%%%%%%%%%%%%%%%
So far we have restricted ourselves to the case of isotropic exchange couplings. Dropping
this restriction we have to study the spin-1 Kondo model \eqref{eq:ham} 
with the generalized exchange interaction
\begin{equation}
\frac{1}{2}\!\!\sum_{\alpha\alpha' k k' \sigma\sigma'}
\bigl[J_0^\perp (S^x \sigma_{\sigma\sigma'}^x+S^y \sigma_{\sigma\sigma'}^y)
+J_0^z\,S^z \sigma_{\sigma\sigma'}^z\bigr]
		c_{\alpha k\sigma}^\dagger c_{\alpha' k'\sigma'}.
		\label{eq:hamanis}
\end{equation}
As explained in detail in Refs.~\onlinecite{Schoeller09,SchoellerReininghaus09} the
RTRG-FS method integrates out the reservoir degrees of freedom in a two-step procedure.
In the first step one removes the symmetric part of the Fermi function in the reservoir 
contractions in a single (discrete) RG step, which yields a well-defined perturbative
expansion in $J_0$ and $1/D$, where $D$ denotes the band width. For the anisotropic 
spin-1/2 Kondo model this step only yields negligible perturbative corrections to the 
Liouvillian and current kernel.
In the second (continuous) step one then resums the logarithmic divergencies by 
integrating out infinitesimal energy shells in the remaining asymmetric part of the reservoir
contractions. This yields the PMS equations \eqref{eq:PMSeq} as well as 
similar RG equations for the Liouvillian and the current kernel. 

However, the situation is completely different for the anisotropic spin-1 Kondo model 
\eqref{eq:hamanis}. The presence of a finite anisotropy
\begin{equation}
	c^2=\big(J^{z}\big)^2-\big(J^{\perp}\big)^2
\end{equation}
produces the term [see Eq.~(74) in Ref.~\onlinecite{SchoellerReininghaus09}]
\begin{equation}
	-\frac\pi4D\bar G_{12}\tilde G_{\bar2\bar1} =-\pi c^2 D L_\Delta\label{eq:anisotropy-term}
\end{equation}
in the Liouvillian. Here $L_\Delta$ denotes the superoperator 
corresponding to the square of the local spin, $L_\Delta\,.=\comm{S^zS^z}{\,.\,}$.
We note that for spin-1/2 this operator vanishes since $S^zS^z=1/4$, thus the term
\eqref{eq:anisotropy-term} does not appear.

Since Eq.~\eqref{eq:anisotropy-term} contains the 
initial band width $D\gg h_0,V$ it will completely dominate the physics of the system. We distinguish 
two cases: (i) For $c^2>0$ the term \eqref{eq:anisotropy-term} is negative, thus locking
the dot in one of the fully polarized states with $M=\pm 1$ and $A=1/\sqrt{6}$. 
As the energy spacing to the other
states on the dot is proportional to $D$ only elastic cotunneling processes are possible
and the current is given by $I=\pi (J_0^z)^2 V$. (ii) For $c^2<0$ the term 
\eqref{eq:anisotropy-term} is positive and the dot magnetization takes the value $M=0$.
Perturbation theory in the exchange couplings then yields $I=0$.

This behavior was previously observed in the anisotropic spin-1 Kondo model at
equilibrium via a PMS analysis~\cite{Konik-02prb2,SchillerDeLeo08} as well as an NRG
calculation~\cite{Zitko-08}. In fact, the term $S^zS^z$ was shown to be a relevant 
perturbation on the grounds of general symmetry 
considerations~\cite{Konik-02prb2,Affleck-92}.

%%%%%%%%%%%%%%%%%%%%%%%%%%%%%%%%%%%%%%%%%%%
\section{Conclusions}\label{sec:conclusion}
%%%%%%%%%%%%%%%%%%%%%%%%%%%%%%%%%%%%%%%%%%%
In conclusion, we have studied non-equilibrium transport through a quantum dot modeled
by a spin-1 Kondo model in a magnetic field. We obtained analytic results for the 
renormalized magnetic field and g-factor, the magnetization and anisotropy as well as
the differential conductance. The latter shows non-monotonic behavior as a function of the 
bias voltage with a pronounced jump at $V=h$ due to the onset of inelastic 
cotunneling processes. In particular, this jump is strongly enhanced by logarithmic 
corrections as was previously observed in transport experiments on carbon nanotube 
quantum dots~\cite{Paaske-06}.

The transport properties of the spin-1/2 Kondo dot have been studied previously using
the same formalism~\cite{SchoellerReininghaus09}. The main difference in the spin-1
model is the appearance of three additional decay rates which govern the relaxation 
of the spin anisotropy $\propto S^zS^z$ on the dot, a quantity which becomes constant 
in the spin-1/2 model. Hence the relaxation dynamics of the magnetization and 
current turns out to be much richer in the spin-1 model. 
Furthermore the decay rates appear as cut-off parameters in the 
logarithmic terms of the perturbative expansions. As a consequence, the presence of 
additional rates leads to an enhanced broadening of the logarithmic corrections of 
stationary quantities like the susceptibility (see Fig.~\ref{fig:sus_limit}) and the differential
conductance (see Fig.~\ref{fig:conductance}), which are thus less pronounced
than in the spin-1/2 model.

Finally we note that as our approach is based on an expansion in the renormalized 
exchange coupling it is not possible to derive reliable results in the low-energy 
regime $\Lambda_c=\max\{V,h\}< T_K$ where the differences between the fully
screened spin-1/2 model and the underscreened spin-1 model are most 
pronounced. As the underscreened Kondo model can be mapped to an effective
ferromagnetic Kondo model, a perturbative study of the transport properties in this 
regime might be possible using a Majorana diagramatic theory as recently performed 
for the investigation of the magnetotransport~\cite{CanoFlorens}.

%%%%%%%%%%%%%%%%%%%%%%%%%%%%%%%%%%%%%%%%%%%
\acknowledgements
We thank Theo Costi, Herbert Schoeller, and Maarten Wegewijs for very valuable
discussions. We further thank Serge Florens, Dante Kennes, Verena Koerting, 
Sarah M\"uller, and Frank Reininghaus for useful comments.
This work was supported by the German Research Foundation (DFG) 
through FG 723, FG 912, and the Emmy-Noether Program (C.B.M.H. and D.S.).
%%%%%%%%%%%%%%%%%%%%%%%%%%%%%%%%%%%%%%%%%%%

\appendix
\section{Liouvillian}\label{app:liouvillian}
%%%%%%%%%%%%%%%%%%%%%%%%%%%%%%%%%%%%%%%%%%%
Here we present the main modifications required to generalize the
calculations of Ref.~\onlinecite{SchoellerReininghaus09} to the spin-1 Kondo model.
Further details can be found in Ref.~\onlinecite{Hoerig11}.

\emph{Superoperators.}---The basis for the local Liouville space of the dot can be built
up by the superoperators $\vec{L}^+$ and $\vec{L}^-$ defined by
\begin{equation}
	\vec{L}^+A = \vec{S}A,\quad
	\vec{L}^-A = -A\vec{S}.\label{eq:Lpm}
\end{equation}
An explicit representation is given by
\begin{equation}
	\vec{L}^+=\vec{S}\otimes\mathbb{1},\quad \vec{L}^{-}=
	-\vec\zeta\,\mathbb{1}\otimes \vec{S}
\end{equation}
where $\vec\zeta=(1,-1,1)$. In general the action of a superoperator $O$ on an 
arbitrary operator $B$ (acting itself in the Hilbert space) can be written in a matrix notation 
as
\begin{equation}
	(OB)_{ss'}=O_{ss',\bar{s}\bar{s}'}B_{\bar{s}\bar{s}'}.
\end{equation}

\emph{Symmetries.}---The effective dot Liouvillian obeys the same symmetries as the 
Hamiltonian: hermiticity relation, spin-conservation, and probability conservation
\begin{eqnarray}
	&&H=H^\dagger  \Leftrightarrow \left(L_S^\textit{eff}(z)\right)^c = -L_S^\textit{eff}(-z^*),
	\label{eq:hermit}\\
	&&s-s\neq\bar s- \bar s' \Rightarrow \left(L_S^\textit{eff}\right)_{ss',\bar s\bar s'}=0,
	\label{eq:spincon}\\
	&&\tr_S\Big(\rho_S(t)-\rho_S(t_0)\Big)=0 \Leftrightarrow \tr_SL_S^\textit{eff}(z)=0.
	\label{eq:probcon}
\end{eqnarray}
Here the c-transform is defined by
\begin{equation}
	\left(A^c\right)_{ss',\bar s \bar s'} = A^*_{s's,\bar s' \bar s}\label{eq:c-transform}
\end{equation}
and the trace over the system states is given by
\begin{equation}
	\bigl(\tr_SL\bigr)_{\bar{s}\bar{s}'}=\sum_sL_{ss,\bar s\bar s'}.
\end{equation}
If we choose an ordered basis in the Liouville space of the dot by
\begin{equation}
	\lbrace\ket{11},\ket{00},\ket{\text{-}1\text{-}1},\ket{10},\ket{0\text{-}1},\ket{01},\ket{\text{-}10},\ket{1\text{-}1},\ket{\text{-}11}\rbrace,\label{eq:representation}
\end{equation}
the relations \eqref{eq:spincon} leads to a decomposition of 
$L_S^\textit{eff}$ into one $3\times 3$ block with $|s-s'|=0$, two $2\times 2$ blocks with 
$|s-s'|=1$, and two $1\times 1$ blocks with $|s-s'|=2$. The restrictions \eqref{eq:hermit} and
\eqref{eq:probcon} then yield further relations~\cite{Hoerig11} 
between the matrix elements in these blocks.

For vanishing magnetic field $h_0=0$ the model is fully rotational invariant and the dot 
Liouvillian decomposes as \eqref{eq:Leffh0}. The projectors on the different 
spin-subspaces are explicitly given by
\begin{eqnarray}
	L^{s}
		&=&
			-\frac{1}{3}\mathbb{1} +\frac{1}{3}(\vec L^+ \cdot \vec L^-)^2,\\
	L^{t}
		&=&
			\mathbb{1}- \frac{1}{2}\vec L^+ \cdot \vec L^- -\frac{1}{2}(\vec L^+ 
			\cdot \vec L^-)^2,\\
	L^{q}
		&=&
			\frac{1}{3}\mathbb{1}+ \frac{1}{2}\vec L^+ \cdot\vec L^-
			+\frac{1}{6}(\vec L^+ \cdot \vec L^-)^2.
\end{eqnarray}

\emph{Poles of the resolvent.}---The poles of the resolvent $1/[z-L_S^\textit{eff}(z)]$ can be 
parametrized by 
\begin{eqnarray}
	z_1	=&	0,\label{eq:z1}\\
	z_2 =&	-z_2^*						&\equiv -i\Gamma_t^0,\label{eq:z2}\\
	z_3 =&	-z_3^*						&\equiv -i\Gamma_q^0,\\
	z_{4\slash6}	=&-z_{6\slash4}^*	&\equiv \pm h-i\Gamma_t^1,\\
	z_{5\slash7}	=&-z_{7\slash5}^*	&\equiv \pm h-i\Gamma_q^1,\\
	z_{8\slash9} 	=& -z_{9\slash8}^*	&\equiv \pm 2h-i\Gamma_q^2,\label{eq:z89}
\end{eqnarray}
where the real parts are related to the renormalized magnetic field $h$ discussed in 
Sec.~\ref{sec:magnetic-field} while the imaginary parts yield the decay rates 
discussed in Sec.~\ref{sec:rates}. 

%%%%%%%%%%%%%%%%%%%%%%%%%%%%%%%%%%%%%%%%%%%
\section{Evaluation of RG equations}\label{app:rg-calculation}
%%%%%%%%%%%%%%%%%%%%%%%%%%%%%%%%%%%%%%%%%%%
The derivation of the Liouvillian and current kernel for the spin-1 Kondo model 
starts with the results (184)--(187) of Ref.~\onlinecite{SchoellerReininghaus09}.
For the explicit evaluation we decompose the effective dot Liouvillian as
$L_S^\textit{eff}(z)=\sum_i\lambda_i(z)P_i(z)$,
where $\lambda_i(z)$ denote the eigenvalues and the corresponding projectors 
$P_i(z)$'s are determined by the
right and left eigenvectors~\cite{Schoeller09}.
The decomposition yields for the resolvents appearing in the perturbative expansions
\begin{equation}
	\frac{1}{z-L_S^\textit{eff}(z)}=\sum_i \frac{1}{z-\lambda_i(z)}P_i(z) \approx 
	\sum_i \frac{1}{z-z_i}P_i(z_i),\label{eq:app1}
\end{equation}
where we have expanded the eigenvalues and eigenvectors around the poles $z=z_i$.
Eqs. (184)--(187) of Ref.~\onlinecite{SchoellerReininghaus09} can now be evaluated 
by inserting \eqref{eq:app1} and performing the necessary algebra in Liouville space. 
For example in $O(J_c^2)$ logarithmic terms of the form 
\begin{equation}
	G\sum_{i=1}^9 \big(z-z_i\big)\ln\frac{\Lambda_c}{-i(z-z_i)}\,P_i(z_i)G 
	\label{eq:rg-equation}
\end{equation}
appear, where $G\propto J_c$ schematically denotes the superoperator corresponding 
to the exchange interaction. We proceed by evaluating each term in \eqref{eq:rg-equation}
separately. The first term with $i=1$ vanishes~\cite{SchoellerReininghaus09,Schoeller09}
due to $P_1(z_1)G=0$. This implies in particular that all remaining logarithmic terms are 
cut-off by a finite decay rate [see \eqref{eq:z2}--\eqref{eq:z89}] and thus the perturbative
expansion of $L_S^\textit{eff}$ is well-defined as long as $J_c\ll 1$.
The terms with $i=8,9$ can be calculated straightforwardly. The
remaining terms are evaluated using the following approximation (which is not
necessary in the spin-1/2 model): Introducing the short-hand notation
\begin{equation}
	\mathcal{H}_i(z)=\big(z-z_i\big)\ln\frac{i\Lambda_c}{z-z_i},
	\label{eq:broadened-function}
\end{equation}
we have to calculate for $i=2,3$ (the remaining pairs $i=4,5$ and $i=6,7$ can be treated
in the same way)
\begin{widetext}
	\begin{eqnarray}
			G\Big\lbrace
			\mathcal{H}_2(z)P_2(z_2)+
			\mathcal{H}_3(z)P_3(z_3)
			\Big\rbrace G &=&
			\frac{1}{2}G\Big\lbrace
			\mathcal{H}_2(z)\bigl[P_2(z_2)+P_3(z_2)\bigr]+
			\mathcal{H}_3(z)\bigl[P_2(z_3)+P_3(z_3)\bigr]\Bigr\}G\nonumber\\*
			& &+
			\frac{1}{2}G\Big\lbrace
			\mathcal{H}_2(z)\bigl[P_2(z_2)-P_3(z_2)\bigr]-
			\mathcal{H}_3(z)\bigl[P_2(z_3)-P_3(z_3)\bigr]\Bigr\}G\nonumber\\*
			& &\hspace{-30mm}=\frac{1}{2}\bigl[\mathcal{H}_2(z)+\mathcal{H}_3(z)\bigr]
			G\,\mathbb{1}^{3\times 3}\,G
			+\frac{1}{2}\bigl[\mathcal{H}_2(z)-\mathcal{H}_3(z)\bigr]
			G\,\bigl[P_2(z_2)-P_3(z_2)\bigr]\,G+O(J_c^3),\label{eq:app2}
	\end{eqnarray}	
\end{widetext}
where we have used $G[P_2(z)+P_3(z)]G=G\sum_{i=1}^3P_i(z)G=
G\mathbb{1}^{3\times 3}\,G$
with $\mathbb{1}^{3\times 3}$ denoting the projector onto the $3\times 3$ subspace 
with $|s-s'|=0$, and $P_i(z_3)=P_i(z_2)+O(J_c)$. Now the first term in \eqref{eq:app2}
yields the logarithmic corrections in $O(J_c^2)$, e.g. the real parts of the analog terms 
with $i=4,5$ and $i=6,7$ yield the logarithmic corrections to the renormalized magnetic
field \eqref{eq:RMF2}. 

On the other hand, the second term in \eqref{eq:app2} does not
lead to logarithmic corrections in $O(J_c^2)$ as
\begin{equation}
\mathcal{H}_2(z)-\mathcal{H}_3(z)\propto \ln\frac{\Gamma_t^0}{\Gamma_q^0}=O(1).
\label{eq:app3}
\end{equation}
The approximation \eqref{eq:app2} is the main technical modification required for the 
treatment of the spin-1 Kondo model. The possibility to neglect the contributions from 
\eqref{eq:app3}
is essential to derive the analytic results presented in the main text. The approximation
\eqref{eq:app2} obviously yields the combination 
$\mathcal{H}_2(z)+\mathcal{H}_3(z)$ (similar for the other pairs $i=4,5$ and $i=6,7$).
We note, however, the direct relations between the functions $\mathcal{H}_i(z)$
(and thus the rates $\Gamma_i$) and the projectors $P_i(z_i)$ are lost.

We stress again that our calculations rely on \eqref{eq:app3}. Thus if \eqref{eq:app2} were to
contain $\mathcal{H}_1(z)$ (as may be the case in the calculation of two-point
correlation functions) the terms of the form \eqref{eq:app3} become divergent and the
approximation \eqref{eq:app2} is no longer applicable.

Finally we note that the imaginary part of \eqref{eq:app2} [originating from the imaginary part
of $\mathcal{H}_i(z)$] does not contain any logarithmic corrections. Thus the logarithmic
corrections to the imaginary part of $L_S^\textit{eff}(z)$, which yield the logarithmic 
corrections to the decay rates and stationary quantities, only appear in $O(G^3)$. They 
can be evaluated by using a relation similar to \eqref{eq:app2}.

%%%%%%%%%%%%%%%%%%%%%%%%%%%%%%%%%%%%%%%%%%%
\section{Self-consistency equations for decay rates}\label{app:self-consistent}
%%%%%%%%%%%%%%%%%%%%%%%%%%%%%%%%%%%%%%%%%%%
From the poles $z_i$ of the resolvent [see \eqref{eq:z2}--\eqref{eq:z89}] we obtain the 
following self-consistency equations for the decay rates (see Fig.~\ref{fig:rates})
\begin{eqnarray}
	\Gamma_{t/q}^0
		&=&
			2C_1\mp \sqrt{(C_1+C_2)(C_1-C_2)},\label{eq:gammaTQ0}\\
	\Gamma_{t/q}^1
		&=&
			C_3+\frac12 C_4+C_5
			\mp \frac{1}{2} \sqrt{(C_5+2C_3)^2-3(C_2)^2},\qquad
			\label{eq:gammaTQ1}\\
	\Gamma_{q}^2
		&=&4C_3+C_4,\label{eq:gammaQ2}
\end{eqnarray}
where the right-hand side (RHS) of \eqref{eq:gammaTQ0} has to be evaluated at $z=z_{2/3}$, 
the RHS of \eqref{eq:gammaTQ1} at $z=z_{4/5}$, and the RHS of \eqref{eq:gammaQ2}
at $z=z_8$, respectively, and the appearing expressions read
\begin{eqnarray*}
	C_1(z)&=&\frac\pi8J_c^2\Bigl(2\abs{z-h}_1+2\abs{z+h}_1+\abs{z-h+V}_1\\*
	&&+\abs{z+h+V}_1+\abs{z-h-V}_1+\abs{z+h-V}_1\Bigr),\\
	C_2&=&2\pi J_c^2 h,\\
	C_3(z)&=&\frac\pi4J_c^2\Bigr(2\abs{z-2h}_2+\abs{z+V-2h}_2+\abs{z-V-2h}_2\Bigl),\\
	C_4(z)&=&\frac\pi4J_c^2\Bigr(2\abs{z-h}_1+\abs{z+V-h}_1+\abs{z-V-h}_1\Bigl),\\
	C_5(z)&=&\frac\pi4J_c^2\Bigr(2\abs{z}_0+\abs{z+V}_0+\abs{z-V}_0\Bigl).
\end{eqnarray*}
The absolute values are broadened by the decay rates and explicitly given by 
\eqref{eq:abs1}	 as well as
\begin{eqnarray}
	\abs{x}_{0}
		&=&
			\frac2\pi x \left(\arctan\frac{x}{\Gamma_{t\!\!\!\phantom{q}}^{0}}
			+\arctan\frac{x}{\Gamma_{q}^{0}}\right),\\
	\abs{x}_2
		&=&
			\frac2\pi x \arctan\frac{x}{\Gamma_{q}^2}.
\end{eqnarray}

%%%%%%%%%%%%%%%%%%%%%%%%%%%%%%%%%%%%%%%%%%%

\end{document}